\documentclass[aps,pra,superscriptaddress,reprint]{revtex4-2}
\usepackage{amsmath}
\usepackage{amssymb}
\usepackage{mathrsfs}
\usepackage{graphicx}
\usepackage{booktabs}
\usepackage{siunitx}
\usepackage{multirow}
\usepackage[colorlinks=true, letterpaper=true, pdfstartview=FitV, linkcolor=blue, citecolor=blue, urlcolor=blue]{hyperref}

\begin{document}

\title{Sensing Based on Quantum Correlation of Photons in the Weak Nonlinear Regime}
\author{Zi-Qiang Yin}
\affiliation{Key Laboratory of Low-Dimensional Quantum Structures and
		Quantum Control of Ministry of Education, Key Laboratory for Matter
		Microstructure and Function of Hunan Province, Department of Physics and
		Synergetic Innovation Center for Quantum Effects and Applications, Hunan
		Normal University, Changsha 410081, China}
\author{Zhi-Hao Liu}
\affiliation{Key Laboratory of Low-Dimensional Quantum Structures and
Quantum Control of Ministry of Education, Key Laboratory for Matter
Microstructure and Function of Hunan Province, Department of Physics and
Synergetic Innovation Center for Quantum Effects and Applications, Hunan
Normal University, Changsha 410081, China}
\author{Jian Tang}
\affiliation{Key Laboratory of Low-Dimensional Quantum Structures and
Quantum Control of Ministry of Education, Key Laboratory for Matter
Microstructure and Function of Hunan Province, Department of Physics and
Synergetic Innovation Center for Quantum Effects and Applications, Hunan
Normal University, Changsha 410081, China}
\author{Hui Jing}
\affiliation{Key Laboratory of Low-Dimensional Quantum Structures and
Quantum Control of Ministry of Education, Key Laboratory for Matter
Microstructure and Function of Hunan Province, Department of Physics and
Synergetic Innovation Center for Quantum Effects and Applications, Hunan
Normal University, Changsha 410081, China} \affiliation{Institute of
Interdisciplinary Studies, Hunan Normal University, Changsha, 410081, China}
\author{Xun-Wei Xu}
\email{xwxu@hunnu.edu.cn}
\affiliation{Key Laboratory of Low-Dimensional Quantum Structures and
Quantum Control of Ministry of Education, Key Laboratory for Matter
Microstructure and Function of Hunan Province, Department of Physics and
Synergetic Innovation Center for Quantum Effects and Applications, Hunan
Normal University, Changsha 410081, China} \affiliation{Institute of
Interdisciplinary Studies, Hunan Normal University, Changsha, 410081, China}
\date{\today}

\begin{abstract}
Quantum correlation of photons based on quantum interference, such as unconventional photon blockade (UPB), has been extensively studied for realizing single-photon sources in weak nonlinear regime. 
However, how to use this effect for other practical applications is rarely studied.
Here, we propose schemes to realize sensitive sensing by the quantum correlation of photons based on quantum interference.
We demonstrate that UPB can be observed in the mixing field output from a Mach–Zehnder interferometer (MZI) with two cavities in the two arms based on quantum interference.
We show that the second-order correlation function of the output field is sensitive to the parameters of system, and propose schemes to realize angular velocity and temperature sensing by measuring the second-order correlation of the photons output from the MZI.
We find that the second-order correlation function of the output field is much more sensitive to the parameters of system than the mean photon number, 
which provides an application scenario for the quantum correlation of photons in sensitive sensing.

\end{abstract}

\maketitle

\section{Introduction}



Quantum correlation of photons can be described by the correlation functions~\cite{Glauber1963a,Glauber1963b}, which provide a quantitative quantity to characterize the quantum nature of lights.
To date, the application of quantum photon correlation has been expanded from characterizing single photons~\cite{Buckley_2012,RevModPhys.87.347} to demonstrating photonic quantum logic gate~\cite{ShiS2022NatCo,LiM2020PRAPP} and fractional quantum Hall state~\cite{WangC2024Sci}.

The quantum correlation of photons can be manipulated when they passing through an optical system containing strong photon-photon interaction. 
If the probabilities of multiphoton states are suppressed by the nonlinear interaction, then the output field becomes sub-Poissonian, which is referred as conventional photon blockade (CPB)~\cite{PhysRevLett.79.1467}. CPB
has been proposed in various optical platforms~\cite%
{Ridolfo2012PRL,LiuYX2014PRA,Adam2014PRA,PhysRevLett.107.063601,PhysRevLett.107.063602,LiaoJQ2013PRA,XieH2016PRA,Majumdar2013PRB,Huang2018PRL,Huang2022LPRv,Chakram2022NatPh,Zhou2020PRA}%
, and has been observed in the resonators strongly coupled to quantum emitters~\cite%
{Birnbaum2005Natur,Dayan2008Sci,Aoki2009PRL,Faraon2008NatPh,Reinhard2012NaPho,LangC2011PRL,Hoffman2011PRL}.

In the past decade, some novel mechanisms are proposed to manipulate the correlation properties of photons under weak nonlinear interactions~\cite{Andrew2021sciadv,MaYX2023PRA,ZhouYH2022PRAPP,SuX2022PRA,Ben-Asher2023PRL,ZuoYL2022PRA,PhysRevA.93.043857,PhysRevA.90.013839,PhysRevA.90.063805,PhysRevB.97.241301,PhysRevLett.123.013602,PhysRevLett.129.043601,Stefanatos2020PRA}. The one that attracts the most
attention is the unconventional photon blockade (UPB)~\cite{PhysRevLett.104.183601,PhysRevA.83.021802} that the probabilities of
two-photon excitation is suppressed by destructive interference between different transition paths. UPB has been extensively studied in various systems, such as coupled
optomechanical systems~\cite{Xu_2013,savona2013UPB,ZhangWZ2015PRA}, coupled
cavities with second or third order nonlinearities~\cite%
{Ferretti_2013,PhysRevA.88.033836,Gerace2014PRA,PhysRevA.90.043822,PhysRevA.90.033809,Lemonde2014PRA,PhysRevA.91.063808,ZhouYH2015PRA,PhysRevA.96.053810,PhysRevA.96.053827,PhysRevA.104.053718,ShenS2019PRA,Zubizarreta2020LPR,Casalengua2020PRA,Wang_2020NJP}%
, cavity embedded with a quantum dot~\cite%
{PhysRevLett.108.183601,ZhangW2014PRA,TangJ2015NatSR,LiangXY2019PRA,PhysRevLett.125.197402}%
, coupled-resonator chain~\cite{WangY2021PRL,li2023enhancement,lu2024chiral}%
, etc. UPB has been experimentally realized with a quantum dot in the semiconductor cavity quantum electrodynamics system~\cite{Snijders2018PRL} and with a SQUID in superconducting resonators~\cite{Vaneph2018PRL}.


In a recent work, we found that the photon correlation of the mixing fields output from a Mach–Zehnder interferometer (MZI), with two cavities in the two arms, can be manipulated by quantum interference~\cite{liuZH2024}.
We uncover a scaling enhancement of photon blockade in the mixing fields output from the MZT.
However, whether strong quantum correlation in the mixing output fields can also be realized in the weak nonlinear regime is still an open question.
Moreover, we know that MZI~\cite{Born_Wolf_2019} is a good platform for the measurement of the relative phase acquired by the beam along the two paths, which offers applications in versatile sensing based on the interference pattern~\cite{Terrel:09,2018NaPho,2019OptLE.117}.
How to use the photon correlation in the mixing fields output from MZI for sensing is another interesting question.

In this paper, we propose to realize sensitive sensing based on the correlation properties of mixing fields output from a MZI, with two cavities in the two arms based on weak nonlinearity.
We find that strong photon blockade can be realized in the weak nonlinear regime by the quantum interference between different paths for two photons states passing through the MZI.
We also show that the second-order correlation of the photons output from the MZI is sensitive to the system parameters, which is the main mechanism for sensing based on quantum correlation.
As examples, we propose schemes to realize angular velocity sensing~\cite{Chow1985RMP,Stedman_1997,LiR2016PRA,Ren:17,Li:17,Satoshi2017PRA,LiK2018SCPMA,DeCarlo:19,LaiYH2019Natur,Hokmabadi2019Natur,Mao_2020,Zhang_2022} and temperature sensing~\cite{Giazotto2006RMP,Dedyulin2022MeScT,Moreva2020PRAPP,Xu2014OE,KLIMOV2018308,Liao2021LSA,DongCH2009APL,LiBBAPL2010,XieD2024PRR,Purdy2015PRA,WangQ2015PRA,Purdy2017Sci,Chowdhury2019QS,Montenegro2020PRR,Galinskiy2020Optica,Singh2020PRL,Shirzad2024PRA,Mehboudi2019JPhA,Potts2020PRAPP,Mok2021CmPhy,ZhangN2022PhRvP,Kuang2023PRA1,Kuang2023PRA2,Aiache2024PRE,ZhangN2024PRA,Aiache2024PRA,Zhang2022npjQI,Mukherjee2019CmPhy} by measuring the second-order correlation of the photons output from the MZI.
Interestingly, the second-order correlation is much more sensitive to the parameters of system than the mean photon number, so that the sensing schemes by detecting second-order correlation may achieve a higher sensitivity than the schemes by measuring the mean photon number.

The paper is organized as follows.  In Sec.~\ref{BIN}, we present the model
of a MZI with two cavities in the two arms.  We show UPB in the output field of the MZI both numerically and analytically in Sec.~\ref{CIN}. 
We demonstrate how to realize versatile sensing by measuring the second-order correlation of the photons output from the MZI in Sec.~\ref{QS}. 
Finally, we summary our main results in Sec.~\ref{Con}.

\section{Physical model}\label{BIN} 

\begin{figure*}[tbp]
\centering
\includegraphics[bb=35 390 536 698, width=16 cm, clip]{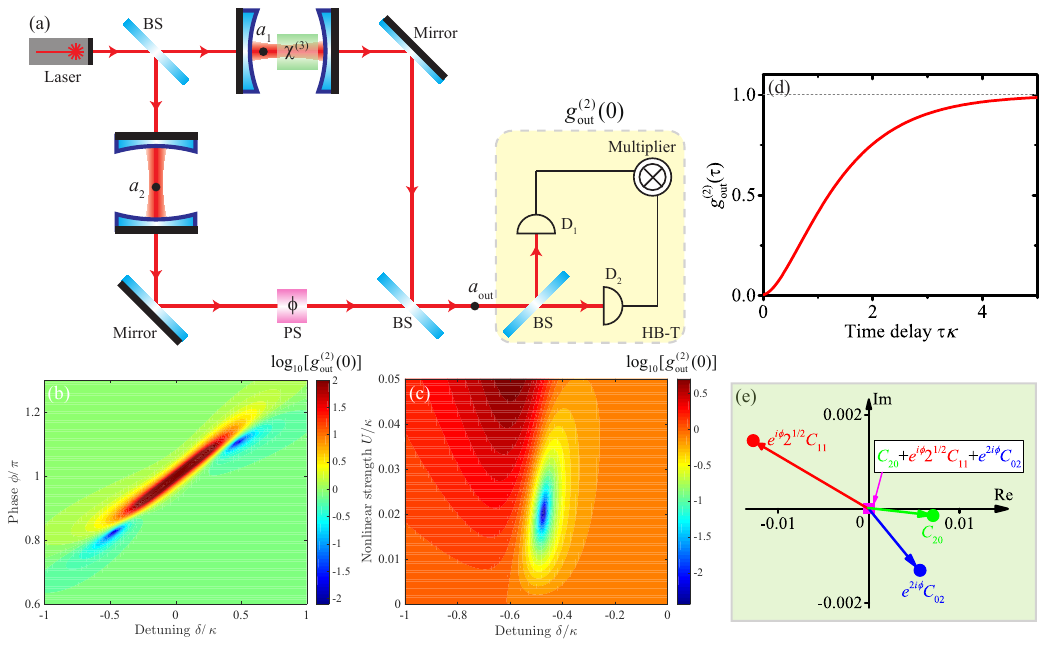}   
\caption{(Color online) (a) A Mach–Zehnder interferometer with two cavities ($a_1$ and $a_2$) in the two arms. A laser is divided into two beams by a $50/50$ beam splitter (BS), and then injected into one of the cavities, respectively. The output fields from these two cavities mix by another $50/50$ BS, a phase shifter (PS) is placed in one of the two paths. The second-order correlation of the output field is measured by a Hanbury-Brown-Twiss (HB-T) set-up. 
The second-order correlation $\log_{10}[g^{(2)}_{\rm out}(0)]$ (b) versus phase $\phi/\pi$ and detuning $\delta/\kappa$ for nonlinear strength $U/\kappa=0.02$, (c) versus $U/\kappa$ and detuning $\delta/\kappa$ for $\phi=0.824\pi$.
(d) The second-order correlation $g^{(2)}_{\rm out}(\tau)$ versus time delay $\tau \kappa$, and (e) the probability coefficients of two photons in the cavities ($C_{20}$, $C_{11}$, and $C_{02}$) and in the output field ($C_{20}+e^{i\phi}\sqrt{2}C_{11}+e^{2i\phi}C_{02}$), for $\delta\approx-0.47\kappa$, $\phi=0.824\pi$, and $U/\kappa=0.02$. 
The parameters are $\varepsilon/\kappa=0.1$ and $\Delta_1=0$.}
\label{fig1}
\end{figure*}

We consider a MZI with two cavities ($a_1$ and $a_2$) in the two arms, as shown in Fig.~\ref{fig1}(a). A laser is divided into two beams by a $50/50$ beam splitter (BS), and then injected into one of the cavities, respectively. The output fields from these two cavities mix by another $50/50$ BS, and a phase shifter (PS) is placed in one of the two paths. The second-order correlation of the mixing output field is measured by a Hanbury-Brown-Twiss (HB-T) set-up. 
In the frame rotating at the probe laser frequency $\omega_p$, the system can be described by the Hamiltonian ($\hbar=1$),
\begin{eqnarray}
H &=&\Delta_1 a_{1}^{\dag }a_{1}+U a_{1}^{\dag }a_{1}^{\dag
}a_{1}a_{1}+i\varepsilon \left( a_{1}^{\dag }-a_{1}\right) \nonumber \\
&&+ \Delta_2 a_{2}^{\dag }a_{2}+i\varepsilon \left(
a_{2}^{\dag }-a_{2}\right) ,
\end{eqnarray}
where $a_{i}$ and $a^{\dag}_{i}$ are the annihilation and creation operators of the cavity mode $i$ with resonance frequency $\omega_i$ ($i=1,2$), $\Delta_i=\omega_i-\omega_p$ is the detuning of the probe laser from the cavity resonance, $\delta=\omega_2-\omega_1$ is the detuning between the two cavity modes, $U$ is the nonlinear interaction strength in cavity $1$, and $\varepsilon$ is the driving strength on each cavity mode.

The two mixing fields $a_{\mathrm{out}}$ and  $A_{\mathrm{out}}$ output from the MZI can be obtained according to the input-output relation~\cite{Gardiner1985PRA}, as
\begin{equation}
a_{\mathrm{out}}=(\sqrt{\kappa _{1}}a_{1}+e^{i\phi }\sqrt{\kappa _{2}}a_{2})/\sqrt{2}-a_{\mathrm{vac}}
\end{equation}
and 
\begin{equation}
A_{\mathrm{out}}=(\sqrt{\kappa _{1}}a_{1}-e^{i\phi }\sqrt{\kappa _{2}}a_{2})/\sqrt{2}-a^{\prime}_{\mathrm{vac}},
\end{equation}
where $\kappa_i$ is the decay rate from one of the cavity mirrors, $\phi$ is the relative phase between the output fields of the two arms (can be tuned by the phase shifter), and $a_{\mathrm{vac}}$ ($a^{\prime}_{\mathrm{vac}}$) is the input vacuum fields. As the output filed $A_{\mathrm{out}}$ can be obtained from $a_{\mathrm{out}}$ just by replacing $\phi$ by $\phi+\pi$, we will only focus on the output field $a_{\mathrm{out}}$ in the following.

The quantum correlation of the output field $a_{\mathrm{out}}$ from the MZI can be described by the second-order correlation function
\begin{eqnarray}
g_{\mathrm{out}}^{\left( 2\right) }\left( \tau\right)  &=&\frac{\left\langle a_{%
\mathrm{out}}^{\dag }a_{\mathrm{out}}^{\dag }\left( \tau\right)a_{\mathrm{out}}\left( \tau\right)a_{\mathrm{out}%
}\right\rangle }{\left\langle a_{\mathrm{out}}^{\dag }a_{\mathrm{out}%
}\right\rangle ^{2}}\\
&=&\sum_{j,k,l,m=1}^{2}e^{i n\phi }\sqrt{\kappa
_{j}\kappa _{k}\kappa _{l}\kappa _{m}}\frac{\left\langle a_{j}^{\dag
}a_{k}^{\dag }\left( \tau\right)a_{l}\left( \tau\right)a_{m}\right\rangle }{4N_{\mathrm{out}}^{2}}\nonumber,
\end{eqnarray}%
where $ n= l+m-j-k$, $2N_{\mathrm{out}}=\kappa _{1}\langle a_{1}^{\dag }a_{1}\rangle
+\kappa _{2}\langle a_{2}^{\dag }a_{2}\rangle +2\sqrt{\kappa
_{1}\kappa _{2}}{\rm Re}( e^{i\phi }\langle a_{1}^{\dag
}a_{2}\rangle ) $, and $\tau$ is the time delay between the two photodetectors ($D_1$ and $D_2$ in the HB-T set-up).
There are cross-correlation between the photons in the two cavities (i.e., $\langle a^{\dagger}_2 a_2 a^{\dagger}_1 a_1\rangle$, $\langle a^{\dagger}_1 a^{\dagger}_1 a_2 a_2\rangle$, $\langle a^{\dagger}_1 a^{\dagger}_1 a_1 a_2\rangle$, and $\langle a^{\dagger}_2 a^{\dagger}_1 a_2 a_2\rangle$), and there are phase factors $e^{in\phi }$ in front of the terms,
which can be negative and lead to the strong photon blockade in weak nonlinear regime based on quantum interference.

The system dynamics is governed by the master equation~\cite{Carmichael1993} for the density matrix $\rho$:
\begin{equation}
\frac{d\rho}{dt}=-i\left[ H,\rho \right] +\sum_{i=1,2}\kappa
_{i}\left( 2a_{i}\rho a_{i}^{\dag }-a_{i}^{\dag }a_{i}\rho -\rho a_{i}^{\dag }a_{i}\right), 
\end{equation}
In the following discussions, we will set $\kappa_1=\kappa_2=\kappa$ for simplicity, and the amplitude of the probe field is weak $\varepsilon=\kappa/10$.

\section{Photon blockade with weak nonlinearity}\label{CIN}

First of all, let us discuss how to realize the photon blockade with weak nonlinearity in the MZI.
The second-order correlation of the output field $\log_{10}[g^{(2)}_{\rm out}(0)]$ is plotted as a function of the phase $\phi/\pi$ and detuning $\delta/\kappa$ for nonlinear strength $U/\kappa=0.02$ in Fig~\ref{fig1}(b). 
Strong photon anti-bunching $g^{(2)}_{\rm out}(0)\ll 1$ is obtained around the parameters ($\delta\approx-0.47\kappa$, $\phi=0.824\pi$) and ($\delta\approx 0.47\kappa$, $\phi=1.104\pi$), with the weak nonlinearity ($U=0.02\kappa$). 
The second-order correlation $\log_{10}[g^{(2)}_{\rm out}(0)]$ is also plotted as a function of the nonlinear interaction strength $U/\kappa$ and detuning $\delta/\kappa$ for $\phi=0.824\pi$ in Fig~\ref{fig1}(c).
The minimal value of $\log_{10}[g^{(2)}_{\rm out}(0)]$ is obtained around $U=0.02\kappa$.
The second-order correlation $g^{(2)}_{\rm out}(\tau)$ versus time delay $\tau \kappa$ is shown in Fig~\ref{fig1}(d).
Different from the rapid oscillations induced by the coupling between two calvities predicted in weakly nonlinear photonic molecules~\cite{PhysRevLett.104.183601,PhysRevA.83.021802}, here, there is no oscillation behavior as the delay
time going on and the scale of the time delay is determined by the decay rate of the cavities as $1/\kappa$.

In order to understand the origin of the strong antibunching, we consider a non-Hermitian Hamiltonian including the decay effect as $H'=H-i\kappa _{1}a^{\dagger}_1a_1-i\kappa _{2}a^{\dagger}_2 a_2$, and use the wave function terminated to the two-photon states as
\begin{eqnarray}
\left\vert \psi \right\rangle  &=&C_{00}\left\vert 0,0\right\rangle
+C_{10}\left\vert 1,0\right\rangle +C_{01}\left\vert 0,1\right\rangle  
\nonumber \\
&&+C_{20}\left\vert 2,0\right\rangle +C_{11}\left\vert 1,1\right\rangle
+C_{02}\left\vert 0,2\right\rangle,
\end{eqnarray}%
for weak probe fields $\varepsilon\ll \{\kappa _{1},\kappa _{2}\}$.
Substituting the non-Hermitian Hamiltonian and wave function into the Schr\"{o}dinger's equation 
$i\hbar \frac{\partial }{\partial t}\left\vert \psi \right\rangle =H^{\prime
}\left\vert \psi \right\rangle $, we get the dynamical equations for the probability coefficients as
\begin{equation}
i\frac{d }{d t}C_{10}=\left( \Delta _{1}-i\kappa _{1}\right)
C_{10}+i\varepsilon C_{00},
\end{equation}%
\begin{equation}
i\frac{d }{d t}C_{01}=\left( \Delta _{2}-i\kappa _{2}\right)
C_{01}+i\varepsilon C_{00},
\end{equation}%
\begin{equation}
i\frac{d }{d t}C_{20}=\left( 2\Delta _{1}+2U-i2\kappa
_{1}\right) C_{20}+i\sqrt{2}\varepsilon C_{10},
\end{equation}%
\begin{equation}
i\frac{d }{d t}C_{02}=\left( 2\Delta _{2}-i2\kappa _{2}\right)
C_{02}+i\sqrt{2}\varepsilon C_{01},
\end{equation}%
\begin{equation}
i\frac{d }{d t}C_{11}=\left( \Delta _{1}+\Delta _{2}-i\kappa
_{1}-i\kappa _{2}\right) C_{11}+i\varepsilon C_{10}+i\varepsilon C_{01}.
\end{equation}%
In the steady state, i.e., $dC_{ij}/dt=0$, we have linear equations for the probability coefficients as
\begin{equation}
0=\left( \Delta _{1}-i\kappa _{1}\right) C_{10}+i\varepsilon C_{00},
\end{equation}%
\begin{equation}
0=\left( \Delta _{2}-i\kappa _{2}\right) C_{01}+i\varepsilon C_{00},
\end{equation}%
\begin{equation}
0=\left( 2\Delta _{1}+2U-i2\kappa _{1}\right) C_{20}+i\sqrt{2}\varepsilon
C_{10},
\end{equation}%
\begin{equation}
0=\left( 2\Delta _{2}-i2\kappa _{2}\right) C_{02}+i\sqrt{2}\varepsilon
C_{01},
\end{equation}%
\begin{equation}
0=\left( \Delta _{1}+\Delta _{2}-i\kappa _{1}-i\kappa _{2}\right)
C_{11}+i\varepsilon C_{10}+i\varepsilon C_{01},
\end{equation}

As $\varepsilon \ll \{\kappa _{1},\kappa _{2}\}$, we assume
that $C_{00}\approx 1\gg \{|C_{10}|,\:|C_{01}|\}\gg\{| C_{20}|,\:|C_{02}|,\:|C_{11}|\}$, and under the conditions $\Delta
_{1}=0$, $\Delta _{2}=\delta $, $\kappa _{1}=\kappa _{2}=\kappa $, we get the expressions of the probability coefficients as
\begin{equation}
C_{10}=\frac{\varepsilon }{\kappa },
\end{equation}%
\begin{equation}
C_{01}=\frac{-i\varepsilon }{\delta -i\kappa }
\end{equation}%
for single-photon states, and
\begin{equation}
C_{20}=\frac{-i}{U-i\kappa }\frac{\varepsilon ^{2}}{\sqrt{2}\kappa },
\end{equation}%
\begin{equation}
C_{02}=-\frac{1}{\sqrt{2}}\frac{\varepsilon ^{2}}{\left( \delta -i\kappa
\right) ^{2}},
\end{equation}%
\begin{equation}
C_{11}=-\frac{i\varepsilon ^{2}}{\kappa \left( \delta -i\kappa \right) }
\end{equation}%
for two-photon states.

Based on the probability coefficients, the output photon intensity $N_{\rm out}/\kappa$ and second-order correlation function $g^{(2)}_{\rm out}(0)$ can be expressed as
\begin{equation}
N_{R,\mathrm{out}}/\kappa\approx \frac{1 }{2}\left\vert C_{10}+e^{i\phi
}C_{01}\right\vert ^{2},
\end{equation}%
and
\begin{equation}
g_{\mathrm{out}}^{\left( 0\right) }\left( 0\right) \approx \frac{2\left\vert
C_{20}+\sqrt{2}e^{i\phi }C_{11}+e^{i2\phi }C_{02}\right\vert ^{2}}{%
\left\vert C_{10}+e^{i\phi }C_{01}\right\vert ^{4}}.
\end{equation}
The optimal conditions for photon blockade is given by
\begin{equation}\label{C_opt}
C_{20}+\sqrt{2}e^{i\phi }C_{11}+e^{i2\phi }C_{02}=0.
\end{equation}%
Under weak nonlinearity $U\ll \kappa $, the optimal conditions are obtianed
as%
\begin{equation}\label{delta_opt}
\frac{\delta_{\rm opt} }{\kappa }\approx \pm \sqrt{\frac{\sqrt{2U/\kappa }-\left(
U/\kappa \right) }{1-\sqrt{2U/\kappa }+\left( U/\kappa \right) }},
\end{equation}%
\begin{equation}\label{phi_opt}
\phi_{\rm opt} =\arg \left[ -\left( 1+i\frac{\delta }{\kappa }\right) \left( 1-e^{i%
\frac{\pi }{4}}\sqrt{\frac{U}{\kappa }}\right) \right] .
\end{equation}
The probability coefficients of two photons in the cavities ($C_{20}$, $C_{11}$, and $C_{02}$) and in the output field ($C_{20}+e^{i\phi}\sqrt{2}C_{11}+e^{2i\phi}C_{02}$) are shown in Fig~\ref{fig1}(e) with the parameters ($\delta\approx-0.47\kappa$, $\phi=0.824\pi$, $U/\kappa=0.02$) for minimal of $g_{\mathrm{out}}^{\left( 0\right) }$, which agree well with Eqs.~(\ref{C_opt})-(\ref{phi_opt}).

\section{Quantum Sensing based on Photon blockade}\label{QS}

\begin{figure}[h]
\includegraphics[bb=29 241 531 628, width=8.5cm, clip]{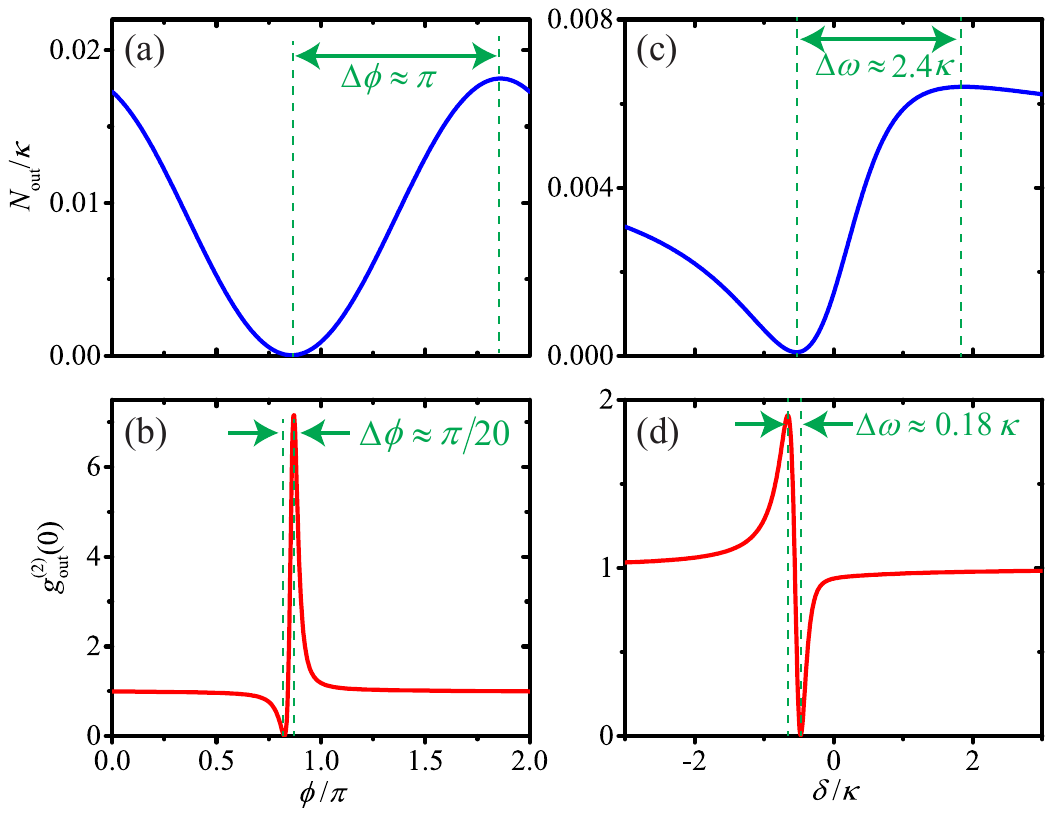}  
\caption{(Color online) The output photon intensity $N_{\rm out}/\kappa$ and second-order correlation function $g^{(2)}_{\rm out}(0)$, (a-b) versus phase $\phi/\pi$ with detuning $\delta/\kappa\approx-0.47$, and (c-d) versus $\delta/\kappa$ with $\phi=0.824\pi$. The other parameters are the same as in Fig.~\protect\ref{fig1}(d).}
\label{fig2}
\end{figure}

We know that both of the output photon intensity $N_{\rm out}/\kappa$ and the second-order correlation function $g^{(2)}_{\rm out}(0)$ depend on the parameters of the system, such as the phase $\phi$ and the detuning $\delta$ as shown in Fig.~\ref{fig2}.
From the figure, we can see that the phase interval $\Delta \phi$ from the maximum to minimum value of the second-order correlation function $g^{(2)}_{\rm out}(0)$ is about $\Delta \phi \approx \pi/20$, which is much narrower than the phase interval ($\Delta \phi \approx \pi$) from the maximum to minimum value of the output photon intensity $N_{\rm out}/\kappa$. The frequency interval $\Delta \omega$ from the maximum to minimum value of $g^{(2)}_{\rm out}(0)$ is about $\Delta \omega \approx 0.18\kappa$, which is also much narrower than the frequency interval ($\Delta \omega \approx 2.4\kappa$) from the maximum to minimum value of $N_{\rm out}/\kappa$.
That is, the second-order correlation function $g^{(2)}_{\rm out}(0)$ is much more sensitive to the parameters than the output photon intensity $N_{\rm out}/\kappa$, which provides us a sensitive quantity for sensing.
As examples, here we will propose two schemes to realize angular velocity sensing and temperature sensing, respectively, by measuring the second-order correlation of the photons output from the MZI.

\subsection{Angular velocity sensing (gyroscope)}

\begin{figure}[h]
\includegraphics[bb=52 43 533 694, width=8.5 cm, clip]{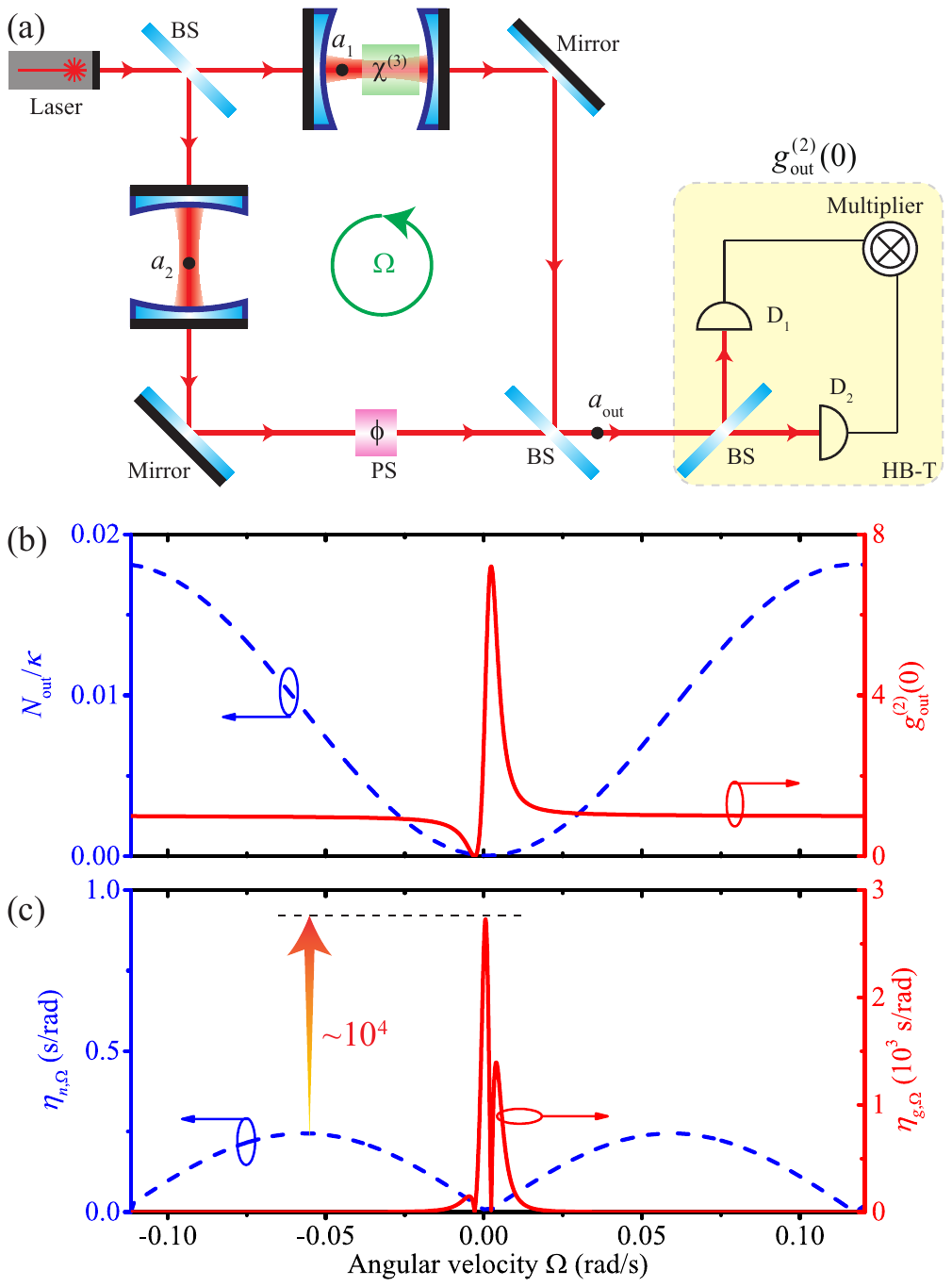}  
\caption{(Color online) (a) The scheme to measure the angular frequency $\Omega$. 
(b) The output photon intensity $N_{\rm out}/\kappa$ and second-order correlation function $g^{(2)}_{\rm out}(0)$ versus $\Omega$.
(c) The sensitivities $\eta_{n,\Omega}$ and $\eta_{g,\Omega}$ versus $\Omega$.
The parameters are the same as in Fig.~\protect\ref{fig1}(d).}
\label{fig3}
\end{figure}

As a specific example, we consider the case that the platform rotates with an angular velocity $\Omega $ as shown in Fig.~\ref{fig3}(a), which leads to an additional phase difference $\Delta \phi$ between the two beams as
\begin{equation}
\Delta \phi =\frac{4\pi A}{\lambda _{0}c}\Omega ,
\end{equation}%
where $A$ is the area enclosed by the optical paths, $\lambda _{0}$ is
wavelength of incident light, and $c$ is the speed of light in vacuum.
We set the phase difference $\phi_0=0.85\pi$ for the platform in the steady state, so we have the total phase difference $\phi=0.85\pi+\Delta \phi$ for the system rotating with an angular velocity $\Omega $.
In the following numerical calculations, we take the parameters~\cite{Culshaw_2006}: $A=10^3\:{\rm m^2}$ and $\lambda _{0}=1550$ nm.

The output photon intensity $N_{\rm out}/\kappa$ and second-order correlation function $g^{(2)}_{\rm out}(0)$ are plotted as functions of the angular velocity $\Omega$ in Fig.~\ref{fig3}(b).
We can see that the second-order correlation function $g^{(2)}_{\rm out}(0)$ is much more sensitive to the change of the angular velocity $\Omega$ than the photon intensity $N_{\rm out}/\kappa$.
To characterize the sensitivity of $N_{\rm out}/\kappa$ and $g^{(2)}_{\rm out}(0)$ to the change of the angular velocity $\Omega$, we can define the sensitivity coefficients as
\begin{equation}
  \eta_{n,\Omega}=\frac{dN_{\rm out}/\kappa}{d \Omega},
\end{equation}
and
\begin{equation}
  \eta_{g,\Omega}=\frac{dg^{(2)}_{\rm out}(0)}{d \Omega}.
\end{equation}
The sensitivities $\eta_{n,\Omega}$ and $\eta_{g,\Omega}$ are plotted as functions of the angular velocity $\Omega$ in  Fig.~\ref{fig3}(c).
The maximal value of $\eta_{g,\Omega}$ is about four orders greater than the one of $\eta_{n,\Omega}$.
It is known that the rotational angular velocity of the earth is about $\Omega_{\rm earth}=7.3\times 10^{-5}$ rad/s, and we have $g^{(2)}_{\rm out}(0)=2.79$ for $\Omega=0$ and $g^{(2)}_{\rm out}(0)=2.97$ for $\Omega=\Omega_{\rm earth}$.
Thus we can detect the rotation of the earth by measuring the second-order correlation function $g^{(2)}_{\rm out}(0)$ of the photons output from the MZI.

\subsection{Temperature sensing}

\begin{figure}[tbp]
\includegraphics[bb=52 54 535 694, width=8 cm, clip]{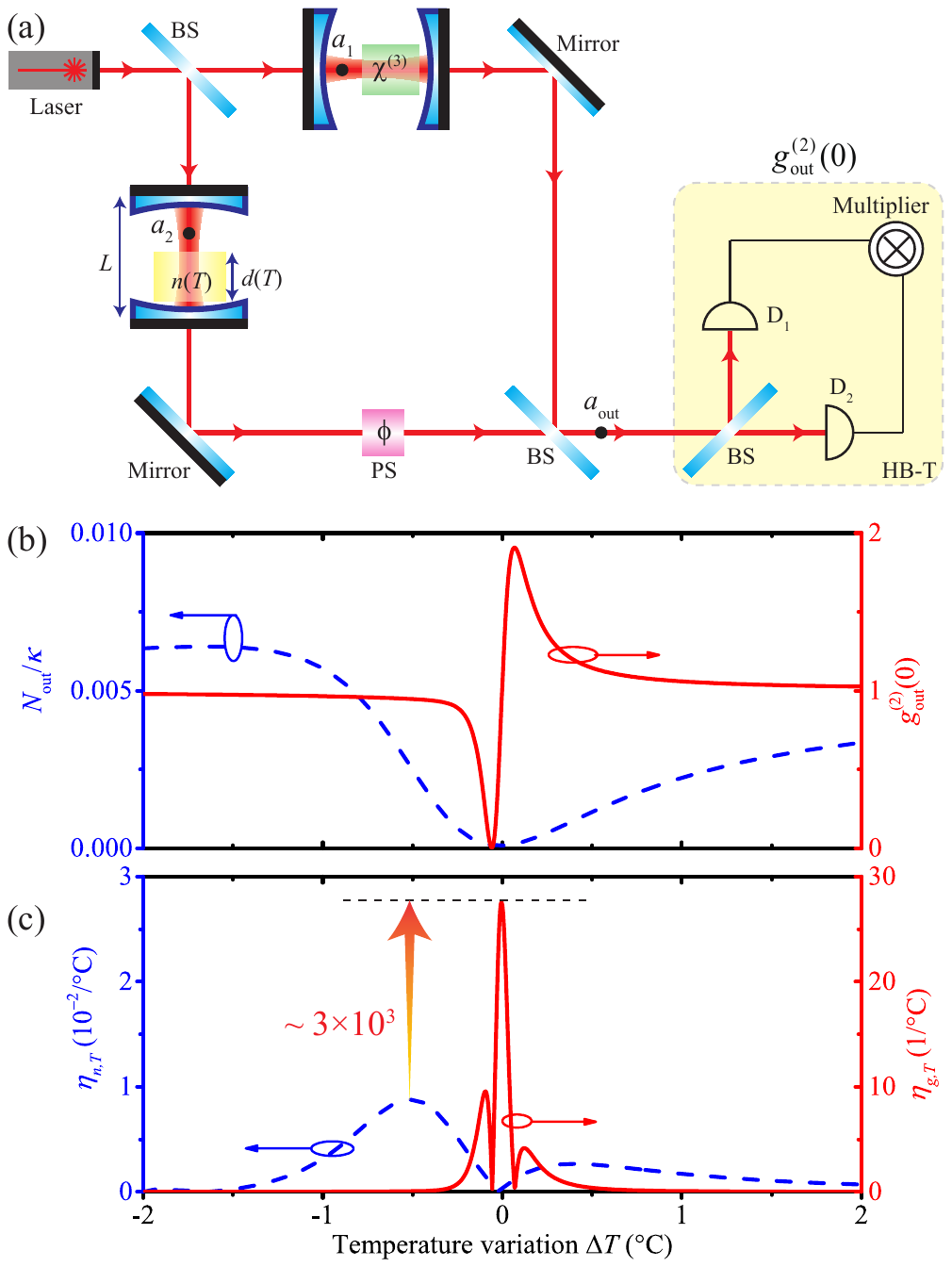}  
\caption{(Color online) (a) The scheme to measure the temperature $\Delta T$. 
(b) The output photon intensity $N_{\rm out}/\kappa$ and second-order correlation function $g^{(2)}_{\rm out}(0)$ versus $\Delta T$.
(c) The sensitivities $\eta_{n,T}$ and $\eta_{g,T}$ versus $\Delta T$.
The parameters are the same as in Fig.~\protect\ref{fig1}(d).}
\label{fig4}
\end{figure}

In this subsection, we consider a dielectric layer in the cavity $a_2$ as shown in Fig.~\ref{fig4}(a), and discuss how to measure the change of the temperature based on the thermal effect.
Due to the thermal expansion, the thickness of the dielectric layer changes as
\begin{equation}
d(T) =d_{0}\left( 1+\alpha \Delta T\right) ,
\end{equation}%
where $d_{0}$ is the thickness of the dielectric layer at a reference temperature, $\alpha$ is the thermal expansion coefficient, and $\Delta T$ is the variation of the temperature.
In addition, as for the thermal-optical effect, the index of refraction also depends on the temperature as
\begin{equation}
n(T) =n_{0}\left( 1+\beta \Delta T\right),
\end{equation}%
where $n_{0}$ is the index of refraction at a reference temperature and $\beta$ is the thermal-optic coefficient.
Based on these two factors, we have the frequency shift of the cavity $a_2$ is given approximately as
\begin{equation}
\delta_{\rm th} \approx -\omega _{2}\left[ \frac{n_{0}d_{0}}{%
L}\beta +\frac{\left( n_{0}-1\right) d_{0}}{L}\alpha \right] \Delta T,
\end{equation}
where $\omega _{2}$ the eigenfrequency of the cavity $a_2$ at a reference temperature with a cavity length $L$.
Under the condition that $\omega _{2}-\omega _{1}=-0.56\kappa $, we have the detuning $\delta=-0.56\kappa+\delta_{\rm th}$.
As a simple example, we consider the dielectric layer as a $\rm SiO_2$ layer, with the typical parameters~\cite{Cavillon:17,Gao18JLT}: $\alpha=5.5\times 10^{-7}/^{\circ}$C, and $\beta=1.0\times 10^{-5}/^{\circ}$C, $n_0=1.45$ at the reference temperature $25^{\circ}$C, $\lambda_0=1550$ nm, $\omega_2= 2\pi c /\lambda_0$, and $\omega_2/\kappa=10^7$. The other parameters are $d_0=0.01$ mm and $L=1$ mm.

The output photon intensity $N_{\rm out}/\kappa$ and second-order correlation function $g^{(2)}_{\rm out}(0)$ are plotted as functions of temperature variation $\Delta T$ in Fig.~\ref{fig4}(b).
We can see that the second-order correlation function $g^{(2)}_{\rm out}(0)$ is much more sensitive to the variation of the temperature  $\Delta T$ than the output photon intensity $N_{\rm out}/\kappa$.
To characterize the sensitivity of $N_{\rm out}/\kappa$ and $g^{(2)}_{\rm out}(0)$ with respect to $\Delta T$, we can define the sensitivity coefficients as
\begin{equation}
  \eta_{n,T}=\frac{dN_{\rm out}/\kappa}{d T},
\end{equation}
and
\begin{equation}
  \eta_{g,T}=\frac{dg^{(2)}_{\rm out}(0)}{d T}.
\end{equation}
The sensitivities $\eta_{n,T}$ and $\eta_{g,T}$ are plotted as functions of the temperature variation $\Delta T$ in  Fig.~\ref{fig4}(c).
The maximal value of $\eta_{g,T}$ is about $28/^{\circ}$C, which is $3\times 10^3$ times greater than the maximal values of $\eta_{n,T}$.

\section{Conclusions}\label{Con}

In conclusion, we have proposed schemes to realize sensitive sensing based on the correlation properties of mixing fields output from a MZI.
We have demonstrated that strong photon blockade can be achieved in the mixing field output form the two cavities in the two arms a MZI in the weak nonlinear regime.
We also have proposed schemes to realize angular velocity and temperature sensing by measuring the second-order correlation of the photons output from the MZI, and shown that the strong photon blockade is much more sensitive to the parameters of system than the mean photon number.
Our work opens a new way for sensitive sensing based on the quantum correlation of photons in the weak nonlinear regime.


\begin{acknowledgments}
X.W.X. is supported by the National Natural Science Foundation of China (NSFC) (Grants No.~12064010, No.~12247105, and No.~12421005),
the science and technology innovation Program of Hunan Province (Grant No.~2022RC1203), 
and Hunan provincial major sci-tech program (Grant No.~2023ZJ1010).
H.J. is supported by the NSFC (Grant No. 11935006, 12421005), the Sci-Tech Innovation Program of Hunan Province (2020RC4047), the National Key R\&D Program (2024YFE0102400), and the Hunan Major Sci-Tech
Program (2023ZJ1010).
\end{acknowledgments}

\bibliography{ref}

\end{document}